# Modeling fractal structure of city-size distributions using correlation function

Yanguang Chen

(Department of Geography, College of Urban and Environmental Sciences, Peking University, 100871, Beijing, China. Email: chenyg@pku.edu.cn)

**Abstract:** Zipf's law is one the most conspicuous empirical facts for cities, however, there is no convincing explanation for the scaling relation between rank and size and its scaling exponent. Based on the idea from general fractals and scaling, this paper proposes a dual competition hypothesis of city develop to explain the value intervals and the special value, 1, of the power exponent. Zipf's law and Pareto's law can be mathematically transformed into one another. Based on the Pareto distribution, a frequency correlation function can be constructed. By scaling analysis and multifractals spectrum, the parameter interval of Pareto exponent is derived as (0.5, 1]; Based on the Zipf distribution, a size correlation function can be built, and it is opposite to the first one. By the second correlation function and multifractals notion, the Pareto exponent interval is derived as [1, 2). Thus the process of urban evolution falls into two effects: one is *Pareto effect* indicating city number increase (external complexity), and the other *Zipf effect* indicating city size growth (internal complexity). Because of struggle of the two effects, the scaling exponent varies from 0.5 to 2; but if the two effects reach equilibrium with each other, the scaling exponent approaches 1. A series of mathematical experiments on hierarchical correlation are employed to verify the models and a conclusion can be drawn that if cities in a given region follow Zipf's law, the frequency and size correlations will follow the scaling law. This theory can be generalized to interpret the inverse power-law distributions in various fields of physical and social sciences.
**Abstract:** Zipf's law; Pareto distribution; rank-size rule; general fractals; multifractals dimension spectrum; scaling analysis; correlation function; city-size distribution

# 1. Introduction

If a region or a sample is large enough to encompass a great many cities, the size distribution of the cities usually follow Zipf's law (Zipf, 1949). Zipf's law for cities is one of the most conspicuous empirical facts in the social sciences generally (Gabaix, 1999a; Gabaix, 1999b). In urban geography, this empirical regularity is known as the rank-size rule (Anderson and Ge, 2005; Berry, 1961; Bettencourt *et al*, 2007; Carroll, 1982; Knox and Marston, 2006; Vining, 1977). Few social science problems have generated more research than the urban rank-size distribution of cities, and numerous models have been proposed to account for variations in rank-size regularity. However, many of the plausible explanations stand in direct contradiction to each other (Carroll, 1982; Córdoba, 2008). For a long time, there is no convincing explanation for the rank-size rule and the scaling exponent value of city rank-size distribution, despite the frequency with which it has been observed (Johnston *et al*, 1994). Today, the rank-size problem seems to be in a dilemma. On the one hand, there are so many theoretical and empirical researches that it seems as if we need no more new models and cases. On the other, the pending problem requires further theoretical study before it will lead us to the underlying rationale of the empirical rule.

In fact, Zipf's law and Pareto distribution are two different sides of the same coin (Gabaix and Ioannides, 2004; Newman, 2005; Reed, 2001). The Pareto distribution is also called Pareto's law since probability distributions are sometimes termed 'laws' (Hardy, 2010). Both Pareto's law and Zipf's law can be associated with fractal distribution (Batty and Longley, 1994; Chen and Zhou, 2003; Frankhauser, 1998; Mandelbrot, 1983). Chen and Zhou (2004) once proposed a dual multifractals model consisting of multi-Pareto-dimension spectrum and multi-Zipf-dimension spectrum to characterize city rank-size distribution. Generally speaking, a multifactals model is always based on generalized correlation function (Feder, 1988; Mandelbrot, 1999; Zhang, 1995). Correlation function is one of the very useful tools in urban studies (Chen, 2009; Chen and Jiang, 2009). If we integrate the idea from multifractals, correlation function, and scaling analysis, we can obtain new insight into the rank-size rule of cities and its scaling exponent.

Recent years, a series of interesting studies on or explanations for the rank-size regularity has been published (e.g. Batty, 2006; Batty, 2008; Blasius and Tönjes, 2009; Ferrer i Cancho and Solé, 2003; Ferrer-i-Cancho and Elvevåg, 2010; Serrano *et al*, 2009; Xu and Harriss, 2010). Especially, the empirical law has been generalized from systems of cities to internal structure of cities as systems, e.g. street hierarchies (Jiang, 2009). These fruits from various fields inspire me to make new researches on city rank-size distribution. This paper will resolve the following problems for the rank-size regularity. First, I construct two correlation function based on Pareto's law and Zipf' law, respectively. By scaling analyses, the value intervals of the scaling exponents of the city rank-size distribution are derived. Second, I present a dual competition hypothesis to explain the scaling exponent values, illuminating why the Pareto exponent approaches 1. Third, mathematical experiments and empirical analysis are performed to verify the theoretical models and inferences. In the context, the scaling exponent includes the Pareto exponent and the Zipf exponent, the former is also called capacity dimension or the zero order correlation dimension, the latter is also termed Zipf dimension, which equals the reciprocal of the Pareto exponent in theory.

## 2. Models

### 2.1 Discrete correlation functions

Suppose there is a region with $N$ cities inside. The size distribution of the $N$ cities follows general Zipf's law, that is

$$P_k = P_1 k^{-d}, \tag{1}$$

where $P_k$ refers to the population of the $k$th city, $P_1$ to the population of the largest city, $k$ to the size rank of the $k$th city in the set, and $d$, the scaling exponent, which also called "Zipf dimension" due to the its association with fractal dimension of urban hierarchy (Chen and Zhou, 2003; Zhang, 1995). Zipf's law suggests a Pareto distribution (Krugman, 1996; Newman, 2005). It is easy to prove that the density function of Pareto distribution is a special density correlation function. Let $f(x)$ represent the number of cities with size over $x$, the discrete correlation function can be defined as

$$C(r) = \frac{1}{N^2} \sum_x f(x) f(x-r), \tag{2}$$

where $x$ is the city size scale, $r$ denotes a "scale displacement factor". Equation (2) means that for the cities with size $x$, what is probability of finding the cities with size "$x$-$r$". Suppose that the city size ($x$) is measured with urban population ($P$), and $f(x)$ is fixed as $f(x)=1$. Numbering the cities as $i, j$ ($i, j=1, 2, …, N$), we can reconstruct the above correlation function by means of Zipf's law and yield

$$C(r) = \frac{1}{N^2} \sum_i^n \sum_j^n H(|P_i - P_j| - r) = \frac{1}{N^2} \sum_{i=1}^n \sum_{j=1}^n H(P_1|i^{-d} - j^{-d}| - r). \tag{3}$$

in which $P_i$, $P_j$ are the size of cities ranked $i$ and $j$, and $H(\cdot)$ denotes Heaviside's function, which can be expressed as

$$H(\cdot) = \begin{cases} 1, & r \leq |P_i - P_j| \\ 0, & r > |P_i - P_j| \end{cases}. \tag{4}$$

This implies that the correlation between two cities is stronger the larger the difference of sizes is. From equation (2) to equation (3), the frequency correlation is replaced by size correlation of cities. The correlation is of scaling invariance if the function follows the power law

$$C(r) = C_1 r^{-D_2}, \tag{5}$$

where $r$ indicates the "yardmeasure" of city size, $D_2$ denotes the correlation dimension of city size distribution, and $C_1$ the proportionality coefficient. For simplicity, we can take $C_1=1$ by normalizing the data. Generally speaking, the correlation dimension in fractal theory implies the second order correlation dimension of general fractals. The mathematical experiments and empirical analysis will be performed by using equations (3) to (5) (see Section 3).

The general correlation function, equation (3), gives a density-density correlation function (the *point-point correlation function*), reflecting the size correlation between any two cities. If we fix one city, say, $P_j$, the density-density correlation function will be reduced to a central correlation function (the *one-point correlation function*). In this instance, all cities are correlated with only one city ($P_j$). Without loss of generality, we may assume $P_j = P_{min}$, where $P_{min}$ denotes the population of the smallest city in the set. Thus we have a central correlation function

$$C(r) = \frac{1}{N} \sum_{i=1}^N \sum_{j=1}^N H(|P_i - P_{min}| - r) = \frac{1}{N} \sum_{i=1}^N H[P_i - (P_{min} + r)]. \tag{6}$$

Rescaling the yardstick as $s = r + P_{min}$ yields

$$C_0(s) = \frac{1}{N} \sum_{i=1}^N H(P_i - s), \tag{7}$$

where $s$ denotes the rescaled yardmeasure, and the Heaviside function should be rewritten as

$$H(\cdot) = \begin{cases} 1, & s \leq P_i \\ 0, & s > P_j \end{cases}. \tag{8}$$

If the central correlation function follows the power law, we deduce the Pareto distribution function such as

$$C_0(s) \propto s^{-D_0}, \tag{9}$$

where $D_0$ refers to the fractal dimension of city-size distribution. This suggests that the Pareto function is a special case of correlation function, and the scaling exponent is in fact the zero order correlation dimension termed "capacity dimension" (Chen and Jiang, 2009; Williams, 1997). By the correlational analysis, Pareto's law and Zipf's law will be integrated into the same framework.

**2.2 Continuous correlation functions based on Pareto's function**

The discrete correlation functions are useful in practice, especially in data fitting/analysis and mathematical experiments. However, it is not easy for us to make theoretical transformation and model deduction. In order to derive new parameter relations, we should substitute the continuous form for the discrete form of mathematical models (Chen, 2010; Casti, 1996). The function of Pareto distribution, equations (7) and (9), can be equivalently re-expressed as

$$N(s) = \sum_{i=1}^{N} H(P_i - s) = N_0 s^{-D_0}, \tag{10}$$

where $N_0=1$ denotes the proportionality coefficient. Thus the density function is

$$\rho(s) = \frac{N(s)}{N} = C_0(s), \tag{11}$$

in which, as indicated above, $N$ denotes the total number of cities in a region. Based on equation (11), we can construct a continuous density-density correlation function such as

$$C(r) = \int_{-\infty}^{\infty} \rho(s)\rho(s-r)\mathrm{d}s = \frac{1}{N^2}\int_{-\infty}^{\infty} N(s)N(s-r)\mathrm{d}s, \tag{12}$$

where $r$ is the scale factor of city size. This can be termed "Pareto correlation function", indicating correlation of city frequency. It is easy to demonstrate that the Pareto correlation function, equation (12), follows the scaling law:

$$\begin{aligned}C(\lambda r) &= \frac{1}{N^2}\int_{-\infty}^{\infty} s^{-D_0}(s-\lambda r)^{-D_0}\mathrm{d}s \\ &= \frac{1}{N^2}\int_{-\infty}^{\infty} (\lambda y)^{-D_0}(\lambda y-\lambda r)^{-D_0}\mathrm{d}(\lambda y), \\ &= \lambda^{1-2D_0}C(r)\end{aligned} \tag{13}$$

where $\lambda$ refers to a scaling factor, and $y=s/\lambda$ to the replacement of $s$. Variable replacement is a very important technique in scaling analysis of mathematical models. Apparently, the solution to the above functional equation is

$$C(r) \propto r^{-(2D_0-1)}. \tag{14}$$

Comparing equation (14) with equation (5) shows the correlation dimension relation as below

$$D_2 = 2D_0 - 1. \tag{15}$$

In theory, a fractal dimension can be treated as a special case of in the spectrum of generalized correlation dimension, namely, a correlation dimension in a broad sense. In equation (15), $D_0$ denotes the zero-order correlation dimension (the moment order equals 0), corresponding to the capacity dimension, while $D_2$ indicates the second order correlation dimension (the moment order is 2), corresponding to the correlation dimension in a narrow sense. The multifractals dimension is a monotonic decreasing quantity with the moment order (Arneodo *et al*, 2008; Chen and Zhou, 2003; Feder, 1988; Mandelbrot, 1999; Stanley and Meakin, 1988). Therefore, $D_0$ is greater than or equal to $D_2$ for ever, that is, $D_0 \geq D_2$. This suggests the first inequation in the form

$$2D_0 - 1 \leq D_0, \tag{16}$$

which implies $D_0 \leq 1$. The numerical relationships between the capacity dimension and the correlation dimension are displayed in Table 1. Obviously, if and only if $D_0 \leq 1$, we will have $D_0 \geq D_2$, and the general fractal dimension spectrum is normal. Otherwise, the multifractals dimension spectrum will fall into disorder. On the other hand, if $D_0 \leq 0.5$, the correlation dimension $D_2 \leq 0$, and this cannot be accept in theory. A conclusion can be drawn that the proper capacity dimension of city-size distributions comes between 0.5 and 1, namely, $0 < D_0 \leq 1$.

**2.3 Continuous correlation functions Based Zipf's law**

The above frequency correlation is based on Pareto's density distribution function. Actually, we can also construct a correlation function based Zipf's law. Generalizing the discrete rank variable ($k$) in equation (1) to a continuous metric variable, we have

$$C(k) = \int_{-\infty}^{\infty} p(k)p(k-l)\mathrm{d}k = \frac{1}{P^2}\int_{-\infty}^{\infty} P(k)P(k-l)\mathrm{d}k, \tag{17}$$

in which $l$ represents a scale factor of city rank, and $P$, the total urban population. This can be termed "Zipf correlation function" indicative of size correlation. For simplicity, we don't change the symbol $k$ for the conversion from discrete distribution to continuous process. By analogy with equation (13), we can prove that the Zipf correlation function, equation (17), satisfies the following scaling relation

$$C(\lambda k) = \frac{1}{P^2} \int_{-\infty}^{\infty} (\lambda z)^{-d_0} (\lambda z - \lambda l)^{-d_0} \, d(\lambda z) = \lambda^{1-2d_0} C(k), \tag{18}$$

where $z=k\lambda$ is the substitute of $k$, and $d_0$ is used to replace $d$ to indicate the zero order Zipf exponent. The solution to equation (18) is a power function as

$$C(k) \propto k^{-(2d_0-1)} = k^{-d_2}. \tag{19}$$

where scaling exponent $d_2$ represents the second order Zipf dimension (Chen and Zhou, 2004), which can be expressed as

$$d_2 = 2d_0 - 1 = \frac{2}{D_0} - 1. \tag{20}$$

It has been proved that the Zipf dimension spectrum is also a monotonic decreasing quantity with the moment order (Chen and Zhou, 2004). Therefore, equation (20) suggests the second inequation such as

$$\frac{2}{D_0} - 1 \leq \frac{1}{D_0}. \tag{21}$$

That is, $D_0 \geq 1$. The numerical relationships between the zero order Zipf dimension and the second order Zipf dimension are listed in Table 1. Apparently, when and only when $D_0 \geq 1$ or $d_0 \leq 1$, we have $d_0 \geq d_2$, and the Zipf dimension spectrum is normal. Otherwise, the Zipf dimension spectrum will fall into confusion. On the other hand, if $d_0 \leq 0.5$ or $D_0 \geq 2$, the second order Zipf dimension $d_2 \leq 0$, and this is meaningless in theory. The conclusion can be reached that the proper Zipf dimension of city-size distributions also falls between 0.5 and 1, namely, $0.5 < d_0 \leq 1$, accordingly, $1 \leq D_0 < 2$. Combining the two inequalities, relations (16) and (21), yields

$$D_0 = \frac{1}{d_0} = 1. \tag{22}$$

This suggests that, in order to satisfy the rationality of Pareto dimension spectrum and Zipf dimension spectrum at the same time, the scaling exponent of the rank-size distribution must be equal to 1 in theory and close to 1 in practice.

**Table 1** The numerical relation between the capacity dimension and the correlation dimension

| Pareto exponent ($D_0$) | Correlation dimension ($D_2$) | Zipf exponent ($d_0$) | Zipf's correlation exponent ($d_2$) |
|---|---|---|---|
| **0.5** | **0** | 2 | 3 |
| **0.6** | **0.2** | 1.667 | 2.333 |
| **0.7** | **0.4** | 1.429 | 1.857 |
| **0.8** | **0.6** | 1.250 | 1.500 |
| **0.9** | **0.8** | 1.111 | 1.222 |
| **1** | **1** | **1** | **1** |
| 1.1 | 1.2 | **0.909** | **0.818** |
| 1.2 | 1.4 | **0.833** | **0.667** |
| 1.3 | 1.6 | **0.769** | **0.538** |
| 1.4 | 1.8 | **0.714** | **0.429** |
| 1.5 | 2 | **0.667** | **0.333** |
| 1.6 | 2.2 | **0.625** | **0.250** |
| 1.7 | 2.4 | **0.588** | **0.176** |
| 1.8 | 2.6 | **0.556** | **0.111** |
| 1.9 | 2.8 | **0.526** | **0.053** |
| 2 | 3 | **0.500** | **0** |

**Note**: The bold denotes the rational intervals of the scaling exponent values.

### 2.4 A dual competition process of city development

City development in a region consists of two major, apparently contradictory, but essentially compatible, processes. One is that cities try to become more and more in number, the other is that each city tries to become larger and larger in size (Steindl, 1968; Vining, 1977). The former is a process of city number increase indicating external complexity of macro level, while the latter is a process of city size growth indicating internal complexity of micro level. The concepts of external and internal complexity came from biology (Barrow, 1995). The former can be termed *Pareto effect*, while the latter, termed *Zipf effect*. The two processes of urban evolution always come into unity of opposites. In theory, the Zipf distribution can be transformed into a self-similar hierarchy, and the competitive relations between city number and city size follows the inverse power law such as (Chen, 2010)

$$N_m = \mu P_m^{-D}, \qquad (23)$$

where *m* is the level order in an urban hierarchy (*m*=1, 2, 3,…), $N_m$ refers to the number of cities in the *m*th level, $P_m$ to the average size of the $N_m$ cities, $\mu$ to the proportionality coefficient, and *D*, to the fractal dimension of the self-similar hierarchy.

Now, a new hypothesis on the dual competition of city development is proposed as follows. If the Pareto effect plays the leading role in evolution of urban systems, the fractal dimension $D_0$ comes between 0.5 and 1, accordingly, the Zipf dimension ranges from 1 to 2. In contrast, if the Zipf effect plays a dominant part in city development, the fractal dimension $D_0$ comes between 1 and 2, and consequently, the Zipf dimension varies from 0.5 to 1. If the two effects reach equilibrium with each other, the scaling exponents $D_0$ or $d_0$ approaches 1. On the other hand, if $D_0 \leq 1$ or $d_0 \geq 1$, the Pareto dimension spectrum is normal, but the Zipf dimension spectrum is abnormal; if $D_0 \geq 1$ or $d_0 \leq 1$, the Zipf dimension spectrum is rational, but the Pareto dimension spectrum is illogical. The composition of forces of the two effects always leads the scaling exponent to the unit: 1. What is more, the positions of the Pareto effect and the Zipf effect can be in exchange with each other. As soon as the scaling exponent go from one extreme to the other (say, from to $D_0>1$ to $D_0<1$), one effect will change to another effect.

## 3. Material and methods

### 3.1 Mathematical experiments

One of the key points in this paper is such a conjecture that if the size distribution of cities follows Zipf's law, the density-density or size-size correlation function will follow the scaling law. This has been theoretically proved by scaling analysis in Subsections 2.2 and 2.3, based on continuous variables of city rank and size. Now, let's make mathematical experiments based on discrete variables to verify the abovementioned judgment. For simplicity, let $N$=500, that is, consider 500 cities in a region. Suppose that all these cities meet the rank-size distribution defined by equation (1). Thus the city sizes can be abstracted as $p$-sequence such as $\{1, 1/2^p, 1/3^p, …, 1/500^p\}$, where $p$ denotes a subset of $d$. The "yardstick" $r$ ranges from 0 to 1 and the step length of yardstick change is taken as $\Delta r$=1/32, that is, $r$=(0), 1/32, 2/32, …, 31/32, (1).

In practice, for simplicity and perspicuity, equation (3) can be replaced by

$$N(r) = \sum_{i}^{n}\sum_{j}^{n} H(|P_i - P_j| - r) = \sum_{i=1}^{n}\sum_{j=1}^{n} H(P_1|i^{-q} - j^{-q}| - r) . \qquad (24)$$

Correspondingly, equation (5) can be rewritten as

$$N(r) = N_1 r^{-D_2} , \qquad (25)$$

where $N_1=C_1N$ denotes a proportionality constant. This is to say, if we substitute correlation number $N(r)$ for correlation density $C(r)$, the scaling exponent will not change (Chen and Jiang, 2009). We can employ some kind of computer software such as Matlab to carry out the mathematical experiments.

The mathematical results shows that the relations between yardstick $r$ and correlation number $N(r)$ follow the scaling law (Figure 1). The scaling exponents give the second correlation dimension $D_2$ values. The zero order correlation dimension, i.e., capacity dimension, can be estimated with equation (15), that is $D_0=(D_2+1)/2$. Changing $p$ value of the $p$-sequence bears an analogy to change the $d$ value in equation (1). The expected capacity dimension is $D_0^*=1/d=1/p$. There are always errors between the theoretical values derived from the mathematical models with continuous variables and the corresponding computational results based on discrete variables from observations or experiments (Chen, 2010). The squared error between computational capacity dimension and expected capacity dimension can be defined as $e^2=(D_0-D_0^*)^2$. Parts of these results from the least square computation are listed in Table 2 for reference.

From the process and results of the mathematical experiments, we can come to the following judgments. First, if $p \leq 0.5$, the size correlation experiments cannot be implemented, or there is no size correlation. This suggests $d_0>0.5$, and thus $D_0=1/d<2$. Second, only when $1 \leq d<2$, that is, $1 \geq D_0>0.5$, we have $D_0 \geq D_2>0$. Otherwise, the multifractals dimension spectrum or the dimension relations will fall into disorder. Third, when $d \approx 1$, and thus $D_0 \approx D_2 \approx 1$, the computation results is most consistent with the theoretical derivation. Actually, when $d_0 \to 1$, or $D_0 \to 1$, the squared error of computational capacity dimension approaches the least. If $d<<1$, the scaling relation tends to be broken down; if $d>>1$, the deviation extent of scaling relation become very large (Figure 1). The conclusions can be drawn as below. First, the value range of the scaling exponent is $0.5<d_0<2$, or $2>D_0>0.5$. Second, the standard rank-size distribution described by the $p$-sequence is a monofractal distribution rather than a multifractal distribution, and thus the expected fractal dimension is $D_0=D_2=1$, and the corresponding Zipf dimension is $d_0=d_2=1$.

Table 2 Partial results of mathematical experiments for size correlation analysis of city rank-size distributions

| $p$ | 0.6 | 0.7 | 0.8 | 0.9 | 1.0 | 1.1 | 1.2 | 1.3 | 1.4 | 1.5 | 2.0 |
|---|---|---|---|---|---|---|---|---|---|---|---|

| $D_2$ | 1.5247 | 1.3693 | 1.251 | 1.1466 | 1.0495 | 0.9618 | 0.8826 | 0.8157 | 0.7473 | 0.6903 | 0.4544 |
| $R^2$ | 0.9727 | 0.9852 | 0.9835 | 0.9777 | 0.9668 | 0.9589 | 0.9489 | 0.9362 | 0.9335 | 0.9184 | 0.8199 |
| $D_0$ | 1.2624 | 1.1847 | 1.1255 | 1.0733 | 1.0248 | 0.9809 | 0.9413 | 0.9079 | 0.8737 | 0.8452 | 0.7272 |
| $D_0^*$ | 1.6667 | 1.4286 | 1.2500 | 1.1111 | 1.0000 | 0.9091 | 0.8333 | 0.7692 | 0.7143 | 0.6667 | 0.5000 |
| $e^2$ | 0.1635 | 0.0595 | 0.0155 | 0.0014 | 0.0006 | 0.0052 | 0.0117 | 0.0192 | 0.0254 | 0.0319 | 0.0516 |

**Note**: $R^2$ denotes the correlation coefficient square, i.e., the goodness of fit.

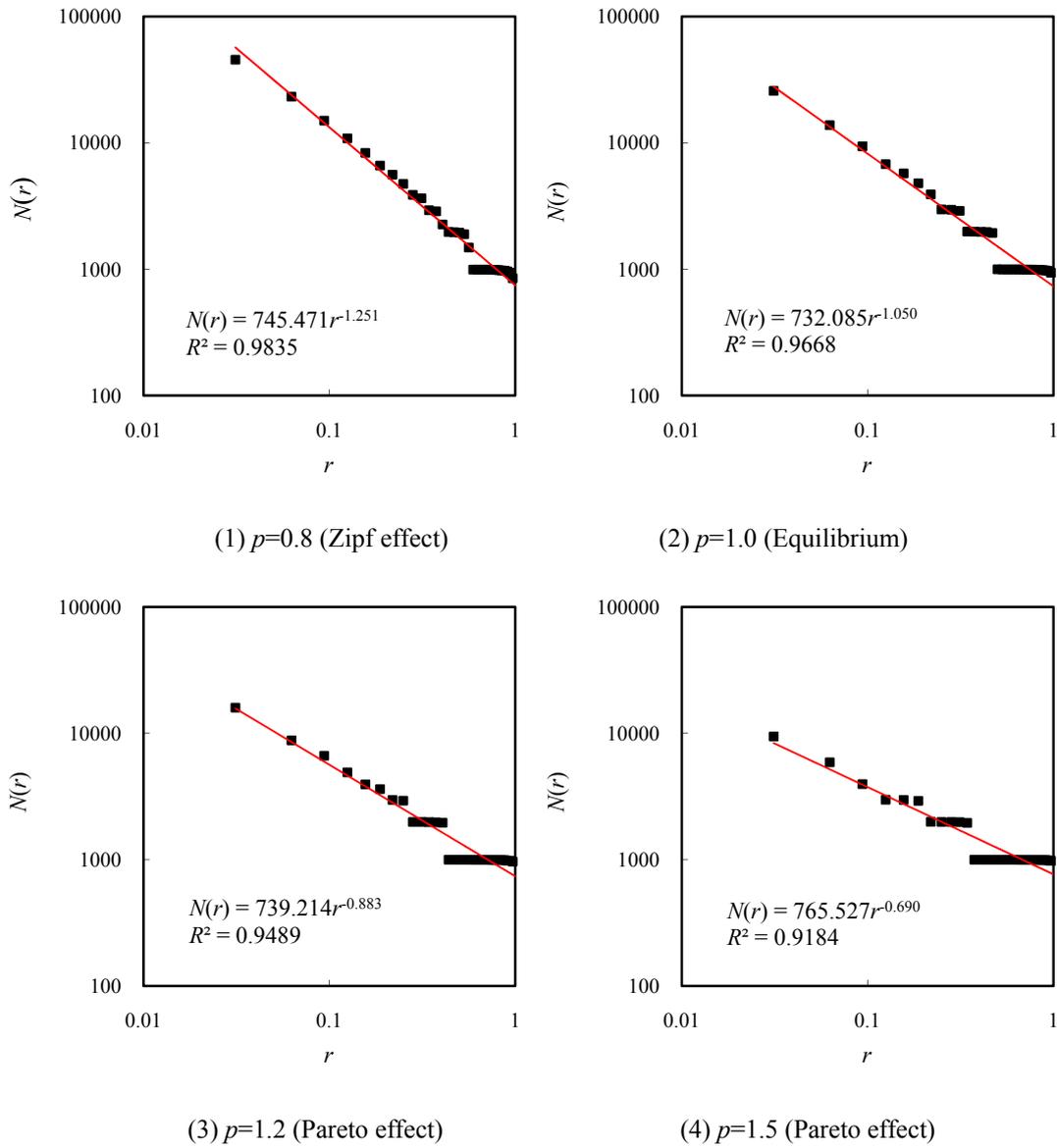

(1) $p=0.8$ (Zipf effect)    (2) $p=1.0$ (Equilibrium)

(3) $p=1.2$ (Pareto effect)    (4) $p=1.5$ (Pareto effect)

**Figure 1** Four typical patterns of size correlation of cities between yardstick and correlation number

## 3.2 Empirical evidences

As an empirical case, the cities of the United States of America (USA) are employed to make a size correlation analysis. The population in urbanized area (UA) is always used to measure the city sizes of America. The 513 largest US cities with UA population over 40,000 according to the 2000

Census are available from internet. These cities comply with Zipf's law in the mass and thus take on a rank-size distribution (Figure 2). On the whole, the correlation function follows the scaling law (Figure 3). Using the least square computation, we can estimate the capacity dimension $D_0$ and the correlation dimension $D_2$. By means of equation (1), the capacity dimension is estimated as $D_0=1/d\approx0.878$; By means of equation (24) and (25), the correlation dimension is estimated as $D_2\approx1.296$. This implies $D_0<D_2$, and the result is abnormal. However, if we replace the least square method with the nonlinear fit method, the results are $D_0=1.225$, $D_2=1.012$, respectively. This time, $D_0>D_2$, and this seems to be normal. As is often the case, different algorithms yield different results and then lead to different conclusions.

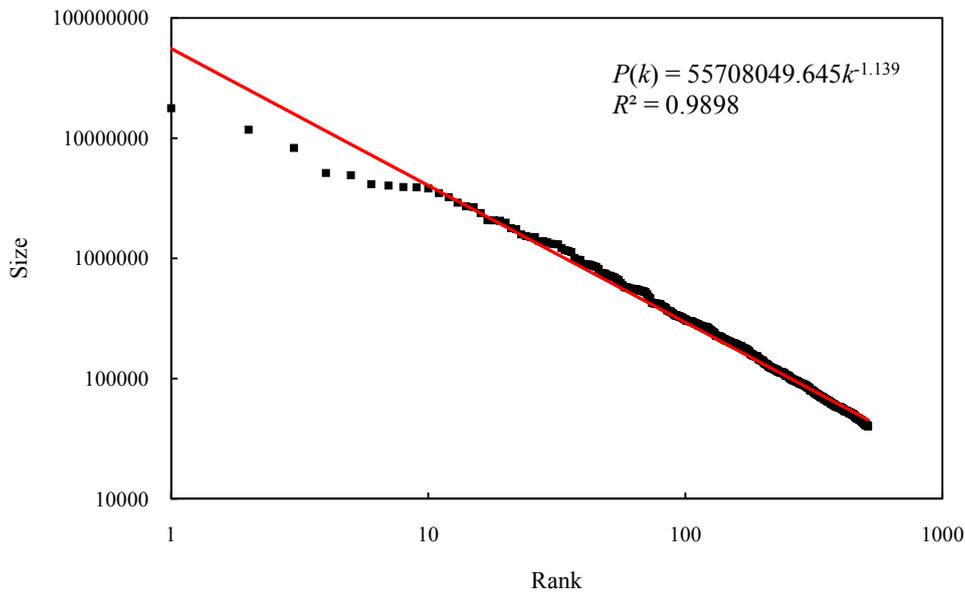

**Figure 2** The rank-size pattern of the first 513 US cities in 2000 (The trend line is given by the least square computation)

(**Data source**: http://en.wikipedia.org/wiki/List_of_United_States_urban_areas)

In fact, the least square method benefits the medium-sized cities and small cities (Figure 2), while the nonlinear fit method favors large cities. This suggests that the large US cities took on Zipf's effect, but the medium and small cities presented the Pareto effect in 2000. The large cities tried to become larger, while the medium and small cities tried to become more and more than ever. However, where statistical average is concerned, the two effects seem to be balanced. The rank-size distribution can be transformed into a self-similar hierarchy, and then we can estimate the fractal dimension of the city-size distribution with the generalized $2^n$ rule (Chen, 2010; Chen and Zhou, 2003). By the $2^n$ rule, the capacity dimension is estimated as $D_0 \approx 0.992$, and $R^2=0.9902$. In this instance, the correlation dimension is expected to approach 1, that is, $D_2 \approx D_0 \approx 1$. The self-similar hierarchy can filter the random disturbance of various noises so that the result is more stable and dependable. It can be seen that the size series of cities in the real world is more complicated than the $p$-sequence in the mathematical world.

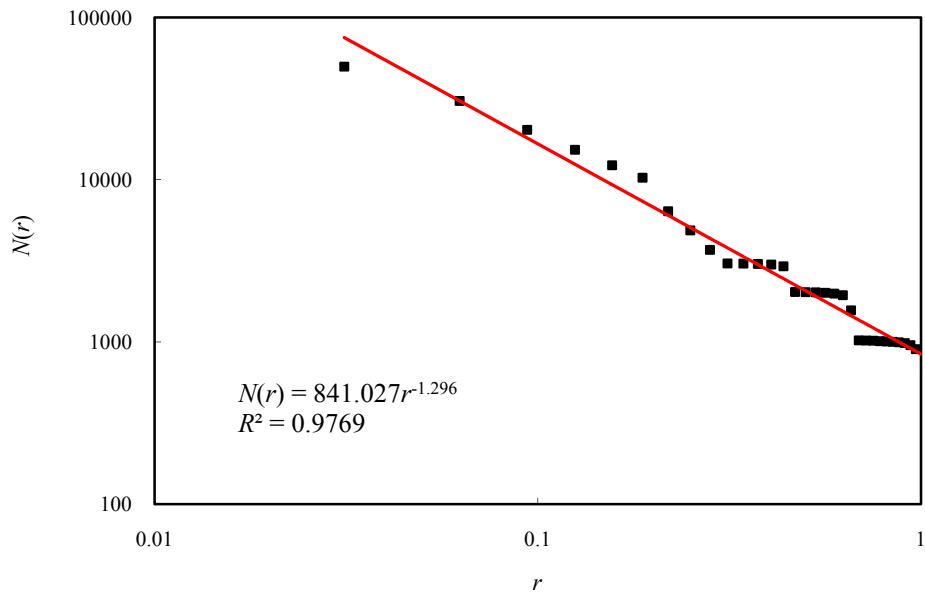

**Figure 3** The point-point size correlation patterns of the US cities based on UA population in 2000

## 4. Discussion and conclusions

Zipf's law is one of the scaling laws in nature and society (Bettencourt *et al*, 2007; Chen, 2010), and scaling laws often typically reflect or even reveal the general principles underlying the structure of a physical problem (West, 2005). The Zipf dimension as well as the Pareto exponent can be treated as a kind of scaling exponent. In order to bring to light the fundament of urban systems, we must estimate the Pareto exponent $D_0$ or Zipf's dimension $d_0$. However, a number of factors affect the parameter estimation. Among these factors, the main are algorithms and city definition. The algorithms in common use include the least square method, the maximum likelihood method, and the nonlinear fit method. The maximum likelihood estimation requires the data meet the normal distribution. Matlab directly provides two algorithms: the least square and nonlinear fit. The former is based on logarithmic scale, while the latter based on conventional scale. If we use the nonlinear fit method, the large cities in the minority will affect the parameter's estimated value; if we use the least square method, the medium and small cities in the majority will influence the result. The essence of algorithmic effect rests with structure of city-size distributions. If the data points distribute along one and only straight line on the log-log plot, the results of parameter estimation from different algorithms should be very close to one another. The US cities seem to form two straight lines rather than single straight line on the logarithmic plot (Figure 2). If we transform the rank-size distribution into a self-similar hierarchy, the problems stemming from algorithms can be resolved to a great extent.

The definition of cities is an important factor impacting the estimation of scaling exponents. In China, there has been no normal or standard definition for cities so far (Jiang and Yao, 2010). In US, there are three basic concepts used to define urban areas and populations, namely, city proper (CP), urban agglomeration or urbanized area (UA), and metropolitan area (MA) (Davis, 1978). The most appropriate one may be UA because it leads to the compatible relations between Zipf's law and the allometric growth law of cities. However, I suggest that the concept of "natural cities" defined by Dr. Jiang and his coworkers should be adopted for urban scaling analysis, because this definition of cities is the most objective one among varied city definitions in use (see Jia and Jiang, 2011; Jiang and Jia, 2011; Jiang B, Liu, 2011). The objectivity of city definition is one of the preconditions of fractal dimension analysis for city rank-size distributions in practice.

If there is no problem in algorithms and urban definition, the scaling exponents, including the Pareto exponent (capacity dimension) and the Zipf dimension, can be employed to make an analysis of city development. The Pareto exponent coming between 0.5 and 1 suggests that the Pareto effect gain an advantage over the Zipf effect, and the cities try to become more in number; The Pareto exponent falling between 1 and 2 implies that the Zipf effect get an advantage of the Pareto effect, and each city tries to become larger (Table 3). The struggle of the two effects leads to two possible results: one is state of equilibrium with scaling exponent equal to 1, and the other is scaling break and data points distribute along two straight lines with different slopes rather than a single line in log-log plots. The scaling pattern of the 513 US cities is actually broken into two lines to some extent, but as a whole, it can be approximately treated as a straight line (Figure 2).

**Table 3** Two effects and two size correlation processes in evolution of urban systems

| Effect | Correlation function | Meaning | Behavior | Equations | Multifractals spectrum | Parameter interval |
|---|---|---|---|---|---|---|
| Pareto effect | Pareto correlation | Frequency correlation | City number increase | (2)-(4), (10)-(12) | General fractal dimension | $0.5<D_0\leq 1$, $1\leq d_0<2$ |
| Zipf effect | Zipf correlation | Size correlation | City size growth | (1), (17) | Zipf dimension | $0.5<d_0\leq 1$, $1\leq D_0<2$ |

**Note**: The general fractal dimension is also called the Pareto-dimension spectrum in the context.

It is impossible to clarify many questions at a time. Some problems remain to be resolved in future. The methods of this study are based on the correlation functions, scaling analysis, and multifractals spectrums. The sum of this paper is as follows. (1) Zipf's law is mathematically equivalent to Pareto's law, but they represent a dual process in urban evolution. Based on Pareto's law, we can construct a frequency correlation function, from which follows a rational value interval of the Pareto exponent as (0.5, 1]; based on Zipf's law, we can construct a size correlation function, from which follows another rational interval of the Pareto exponent as [1, 2). The intersection of the two intervals is $D_0=1/d_0=1$. (2) The dynamical mechanism of city development or urban evolution comes down to two effects. One is the Pareto effect associated with frequency correlation, the other is the Zipf effect associated with size correlation. The two effects are of unity of opposites. If the Pareto effect plays the leading role in urban evolution, cities try to become more and more in number; if the Zipf effect plays a dominant part in city development, each city tries to become larger and larger in size. (3) It is hard for urban evolution to satisfy both sides of the two effects. If the Pareto effect win the advantage over the Zipf effect, the multifractals dimension (multi-Pareto-dimension) spectrum will be rational, but the multi-Zipf-dimension spectrum will be illogical; if the Zipf effect has advantage of the Pareto effect, the multi-Zipf-dimension spectrum will be normal, but the multifractal spectrum will be abnormal. The result of competition of the two effects is either scaling break or the scaling exponent close to 1.

## Acknowledgements:

This research was sponsored by the National Natural Science Foundation of China (Grant No. 40771061. See: https://isis.nsfc.gov.cn/portal/index.asp). The support is gratefully acknowledged.